\def\grl{{\em Geophys. Res. Lett.}}

\def\nu{{\em Nuclear Fusion}}

\documentclass[%
 aip,groupedaddress,
 jmp,%
 amsmath,amssymb,
preprint,
floatfix,
]{revtex4-1}

\usepackage{graphicx}
\usepackage{float}
\usepackage{dcolumn}
\usepackage{bm}
\usepackage{amsmath,amsfonts,amsthm,amssymb}
\usepackage{xcolor}

\begin{document}

\preprint{AIP/123-QED}

\title{ Quantifying the Effect of Non-Larmor Motion of Electrons on the Pressure Tensor}
\author{H. Che\textsuperscript{1,2}, C. Schiff\textsuperscript{2},G. Le\textsuperscript{2}, J. Dorelli\textsuperscript{2}, B. Giles\textsuperscript{2} and T. Moore\textsuperscript{2}}
\affiliation{1: University of Maryland, College Park, MD, 20742, USA }
\affiliation{2: NASA Goddard Space Flight Center, Greenbelt, MD, 20771, USA}%
\date{\today}

\begin{abstract}
In space plasma, various effects of magnetic reconnection and turbulence cause the electron motion to significantly deviate from their Larmor orbits. Collectively these orbits affect the electron velocity distribution function and lead to the appearance of the ``non-gyrotropic" elements in the pressure tensor. Quantification of this effect has important applications in space and laboratory plasma, one of which is tracing the electron diffusion region (EDR) of magnetic reconnection in space observations. Three different measures of agyrotropy of pressure tensor have previously been proposed, namely, $A\varnothing_e$, $D_{ng}$ and $Q$. The multitude of contradictory measures has caused confusion within the community. We revisit the problem by considering the basic properties an agyrotropy measure should have. We show that $A\varnothing_e$, $D_{ng}$ and $Q$ are all defined based on the sum of the principle minors (i.e. the rotation invariant $I_2$) of the pressure tensor. We discuss in detail the problems of $I_2$-based measures and explain why they may produce ambiguous and biased results.  We introduce a new measure $AG$ constructed based on the determinant of the pressure tensor (i.e. the rotation invariant $I_3$)  which does not suffer from the problems of $I_2$-based measures. We compare $AG$ with other measures in 2 and 3-dimension particle-in-cell magnetic reconnection simulations, and show that $AG$ can effectively trace the EDR of reconnection in both Harris and force-free current sheets.  On the other hand, $A\varnothing_e$ does not show prominent peaks in the EDR and part of the separatrix in the force-free reconnection simulations, demonstrating that $A\varnothing_e$ does not measure all the non-gyrotropic effects in this case, and is not suitable for studying magnetic reconnection in more general situations other than Harris sheet reconnection. 

\end{abstract}

\maketitle

\section{Introduction}
\begin{figure}
\includegraphics[scale=0.3, trim=50 125 100 90,clip]{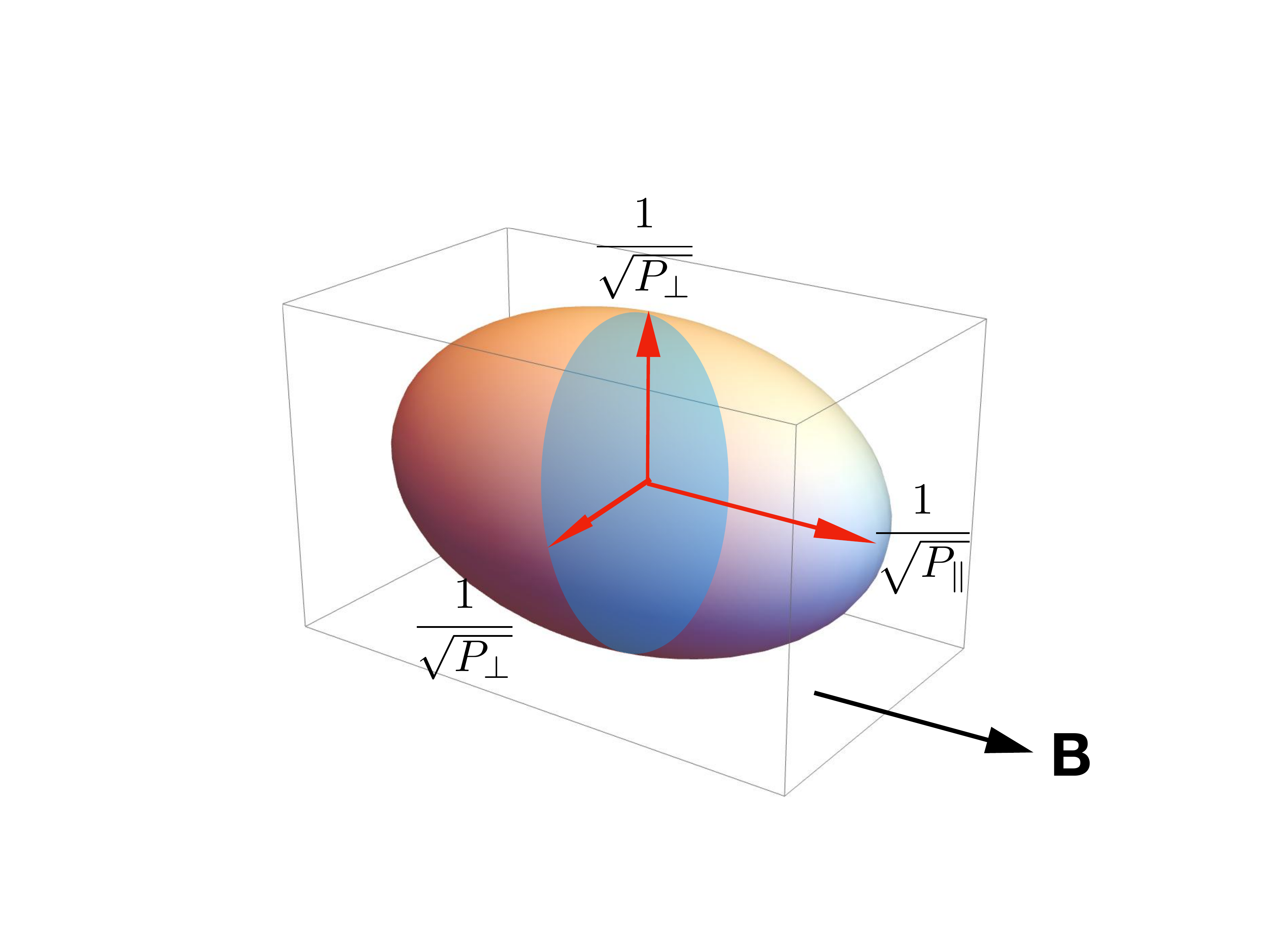}
\caption{The ellipsoid of a gyrotropic pressure tensor in its principle coordinate with one axis parallel to the local magnetic field. The lengths of the three axes of the ellipsoid are $1/\sqrt{P_{\parallel}}$,$1/\sqrt{P_{\perp}}$ and $1/\sqrt{P_{\perp}}$. }
 \label{gfig}
 \end{figure}

The {\it pressure tensor} in kinetic theory \citep{huang66book} is defined as: 
\begin{equation}
P_{ij} = m \int (v_i - \bar{v}_i)(v_j-\bar{v}_j) f(\mathbf{x},\mathbf{v},t) dv^3,
\label{pdefine}
\end{equation}
where ${v}_i$ and ${v}_j$ are the  $i$th and $j$th components of the velocity of a particle and $\bar{v}$ represent the mean velocity of particles. In fluid dynamics the pressure tensor corresponds to the negative of the {\it stress tensor}, but the definition of pressure tensor is more general and does not depend on the validity of the fluid description of plasma. Pressure tensor is real, symmetric, i.e. $P_{ij}=P_{ji}$, and positive semidefinite. In  homogeneous and isotropic plasma, pressure tensor becomes $P_{ij} = P\delta_{ij}$ and is reduced to a scalar. The isotropy can be broken in the presence of a magnetic field.  The motion of charged particles and hence their macroscopic properties may be very different in  directions parallel and perpendicular to the magnetic field. In most situations encountered in space and laboratory plasma, charged particles are magnetized,  i.e., their  gyro-radii being much smaller than the variation scale of the magnetic field, hence their motion can be approximated by the fast cyclotron motion plus a drift. We call such motion {\it Larmor}. Magnetized plasma usually relaxes independently in directions parallel and perpendicular to the magnetic field, and  consequently the velocity distribution function assumes an axisymmetric form, i.e. $f({\bf v}) = f(v_{\parallel},v_{\perp})$.  The pressure tensor is also axisymmetric in the geometric representation\footnote{The contraction of $\mathbb{P}$ yields a scalar quadratic function $f(\mathbf{x}) = \sum x_i P_{ij} x_j$. Since $\mathbb{P}$ is positive semidefinite, the quadratic function defines a family of ellipsoids (including their degenerate forms) with the pressure components as coefficients. Thus the pressure tensor can be conveniently represented by an ellipsoid (or its degenerate forms). In this paper when describing pressure tensor in geometric terms we refer to this representation.}, i.e., having one principle axis of the pressure tensor aligned with the magnetic field as shown in Fig.~\ref{gfig}. The ``gyrotropic" pressure tensor $\mathbb{G}$ with independent $P_{\parallel}$ and $P_{\perp}$ can be written as 
\begin{equation}
\mathbb{G}\equiv\mathbb{P}_{gyro}=\begin{bmatrix}
P_{\parallel} & 0 & 0\\ 0 & P_{\perp} & 0\\ 0 & 0 & P_{\perp}\end{bmatrix}.
\label{gp}
\end{equation}
 
In some important phenomena such as magnetic reconnection in current sheets and turbulence, charged particles encounter large spatial variations of magnetic field or fast changing external forces, and their orbits become far more complicated\cite{zenitani16pop} than the Larmor orbits in uniform or quasi-uniform magnetic fields. These particles can transport momentum between different directions, the collective motion of particles lead to the breaking the axisymmetry or {\it gyrotropy} of the distribution function. 

The simplest manifestation of gyrotropy breaking in pressure tensor is when the perpendicular part of the pressure breaks axisymmetry, but one principle axis of the pressure tensor remains along the magnetic field, thus $\mathbb{P}$ can be written as:
\begin{equation}
\mathbb{P}_{agyro}=\begin{bmatrix}
P_{\parallel} & 0 & 0\\ 0 & P_{\perp 1} & 0\\ 0 & 0 & P_{\perp 2}\end{bmatrix},
\label{ag1}
\end{equation}
where $P_{\perp 1}\neq P_{\perp 2}$ and the description of pressure tensor needs three independent parameters. In general the non-Larmor motion of particles can lead the non-diagonal elements of pressure in Eq. (\ref{ag1}) to become non-zero, and in geometric terms none of principle axes are aligned with the magnetic field: 
\begin{equation}
\mathbb{P}_{agyro}=\begin{bmatrix}
P_{\parallel} & P_{12} & P_{13}\\ P_{12} & P_{\perp 1} & P_{23}\\ P_{13} & P_{23} & P_{\perp 2}\end{bmatrix}.
\label{pp}
\end{equation}

\begin{figure}
\noindent\includegraphics[scale=0.65, trim=20 280 80 100,clip] {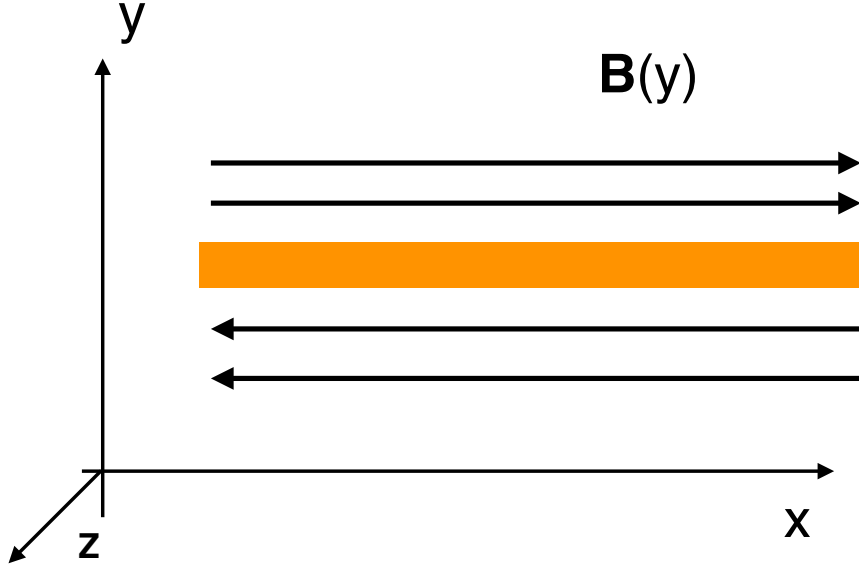}
\caption{An Illustration of current sheet. }
 \label{current}
 \end{figure}

Non-Larmor motion of charged particles can occur in regions with strong inhomogeneity and large shear. A simple example can be found around current density peaks in thin current sheets where the magnetic field reverses its direction \cite{sonnerup71jgr}. In this type of current sheet the magnetic shear is large while the field is weak, a condition that causes the electrons to demagnetize and their orbits are characterized by the so called meandering motion. In an unperturbed (static) current layer with magnetic field configuration shown in Fig.~\ref{current}, the components of the non-relativistic canonical momentum $\mathbf{p}$ of an electron are:
\begin{eqnarray}
p_x=mv_x, \qquad   p_y=mv_y, \qquad \ p_z=mv_z-\frac{e}{c}A_z(y),
\end{eqnarray}
with $\bold{A}$ being the magnetic vector potential, and the Hamiltonian is $H=[p_x^2+p_y^2+(p_z+e A_z/c)^2]/2m$.
The energy of the electron is conserved. The Hamiltonian equations $\partial p_i/\partial t =-\partial H / \partial x_i = 0$ imply that in a static anti-parallel magnetic field, each component of canonical momentum is also conserved.  As electrons move near or cross the magnetic null, the sharp change of magnetic field $B_x=\partial_y A_z(y)$ leads to inhomogeneous drift and meandering motion of particles in the $yz$ directions\cite{vasy75rg}. A consequence of the conservation of canonical momentum and energy is that the change in magnetic momentum is redistributed between the kinetic momentum $m v_y$ and $m v_z$. Collectively the effect causes the electron velocity distribution function to deviate from Maxwellian in directions perpendicular to the magnetic field. The consequence is the nondiagnal elements of the pressure tensor in a field-aligned frame generally become nonzero or the perpendicular elements becomes unequal. An initially gyrotropic pressure tensor eventually becomes 
\begin{equation}
\mathbb{P}=\begin{bmatrix}
P_{\parallel} & 0 & 0\\ 0 & P_{\perp1} & P_{23}\\ 0 & P_{23} & P_{\perp2}\end{bmatrix}
\end{equation}
in any field aligned coordinate except  in the eigenvector aligned  frame, i.e., a frame determined by the magnetic field and the current sheet normal, where the pressure tensor become diagonalized to the form shown in Eq. (\ref{ag1}). It should be noted that the commonly used Harris solution\cite{harris62} of Vlasov equation is obtained under the assumption of global isotropic pressure despite the existence of local meandering motions near the null region. If an out-of-plane electric field exists, the chaotic motions of electrons around null-point become more complex and can produce dramatic non-gyrotropic effects which play an important role in magnetic reconnection\cite{hesse99pop,hesse16grl}.
 
In magnetic reconnection, the situation is more complex than in current sheets associated with anti-parallel magnetic fields. First, the topological change of the magnetic field can break the 2D anti-parallel magnetic field configuration which produces sharply curved field lines and reconnection electric field. In component reconnection, the magnetic field is 3D. The magnetic vector potential is no longer dependent only on the two spatial coordinates perpendicular to the magnetic field. Consequently the Hamiltonian of particles also become functions of all canonical coordinates and no component of canonical momentum is guaranteed to conserve, i.e., $\partial p_i/\partial t \neq 0$\citep{chen93pop,Lit93sp}. More importantly, the motion of particles around the reconnection x-line can become stochastic\citep{speiser65jgr,speiser67jgr,buechner89jgr,chen92jgr}. Such stochastic motion causes diffusion in the velocity space and allows the particles to transfer kinetic momentum between different directions and build up correlations between all velocity components\citep{buechner86jgr}. On the other hand,  kinetic instabilities are common in magnetic reconnection and these instabilities can also produce anisotropic heating and scattering. Electron-scale magnetic reconnection configurations can be produced in turbulence due to anomalous dissipation of magnetic energy\cite{che17pop}. Considering all these complexities, in magnetic reconnection the non-diagonal elements of the pressure tensor in the field-aligned frames are usually non-zero as shown in Eq.~(\ref{pp}), and the principle axis of the pressure tensor also is generally not aligned with magnetic field. 

In space physics, the interest in the non-Larmor effects on pressure tensor originated from an early argument made by Vasyliunas\cite{vasy75rg} that the gradient of non-gyrotropic part of the electron pressure tensor is essential in supporting the reconnection electric field in 2D static collisionless magnetic reconnection. This effect has since been demonstrated in many Harris sheet magnetic reconnection simulations\cite{hesse95jgr, kuznetsova98jgr, pritchett01jgr, ricci04pop, daughton06pop}.  It was then suggested\cite{scudder08jgr,scudder12prl} that this effect may be used  to ``illuminate'' the electron diffusion region (EDR) of magnetic reconnection in {\it in-situ} observations of the magnetosphere, given that other indicators of the EDR are not ubiquitous. In nature, magnetic reconnections generally are not 2D and static, and the pressure gradient is not necessarily the only or the dominant term in the Ohm's law that supports the reconnection electric field. Nevertheless, the singular magnetic field configuration at the vicinity of the x-line is sufficient to produces significant amount of non-Larmor electrons in the EDR\cite{speiser65jgr,speiser67jgr,buechner89jgr,zenitani16pop}. Therefore, the effect of the non-Larmor electrons on the pressure tensor should be significant in most of the reconnection events, not just in 2D simulations, and should be a useful indicator of the EDR.  

In practice, quantification of the non-Larmor effect provides a useful parameter for detection of crossings of the EDR in the time sequence data obtained by space probes, e.g. the recently launched {\it Magnetospheric Multiscale Science} (MMS)\cite{burch16ssr}. While the electron velocity distribution functions  carry detailed information of the statistical properties of the electrons, a quantification of the effects of agyrotropy on the pressure tensor allows easy comparison of observations at different locations and time from space instruments. Such a quantity is particularly  convenient when searching for magnetic reconnection EDRs in large observation datasets, or conducting correlation analysis of pressure tensor and other thermodynamic or electromagnetic properties of plasma.   

Several measures of agyrotropy of pressure tensor have previously been proposed. An early proposal\cite{scudder08jgr}  defines agyrotropy as the relative difference of the two eigenvalues of the perpendicular sub-matrix of the pressure tensor in the field aligned coordinate, i.e. $A\varnothing_e \equiv \mid P_{\perp 1}^{\prime}-P_{\perp 2}^{\prime}\mid/(P_{\perp 1}^{\prime}+P_{\perp 2}^{\prime})$(We ignored the factor of 2 in the original definition.) . In other words, after a rotation around the magnetic field $\mathbb{R}^T\mathbb{P}\mathbb{R}$, here $\mathbb{R}$ is the 2D space rotation matrix, the pressure tensor in Eq.(\ref{pp}) becomes
\begin{equation}
\mathbb{P}=\begin{bmatrix}
P_{\parallel}^{\prime} & P_{12}^{\prime} & P_{13}^{\prime}\\ P_{12}^{\prime} & P_{\perp 1}^{\prime} & 0\\ P_{13}^{\prime} & 0 & P_{\perp 2}^{\prime}\end{bmatrix},
\label{aphi}
\end{equation}
where $\prime$ represents the new values after the rotation.  The definition $A\varnothing_e$ is criticized for not accounting for the contribution from all the non-diagonal components of the pressure tensor\cite{aunai13pop}. $A\varnothing_e$  is clearly defined assuming the deformation of electron orbits only occurs in the perpendicular directions and the momentum transport is constrained to the perpendicular plane (which is not generally the case). In this particular situation the pressure tensor is always field aligned and the definition of agyrotropy as $A\varnothing_e$ is an intuitive choice.  In situations where the deformations of the electron orbits are not confined to the perpendicular directions, as demonstrated in magnetic reconnection simulations\cite{speiser65jgr,speiser67jgr,buechner89jgr,zenitani16pop}, $A\varnothing_e$ clearly misses contributions from orbit distortions that are not confined to the perpendicular plane. To make $A\varnothing_e$ useful, the proponents of  $A\varnothing_e$ have to hope that the distortion of electron orbits in the perpendicular plane always dominates so that $A\varnothing_e$ can still be used as a biased tracer of the EDR. This is a falsifiable proposition, and all we need is an example to show that non-gyrotropic effect can be strong but $A\varnothing_e$ is small, as we will do in \S~\ref{p3d}.

Albeit different measures have been proposed, it is generally agreed that agyrotropy is a scalar that quantifies the departure of the pressure tensor from axisymmetry about the local magnetic field. Mathematically, any pressure tensor $\mathbb{P}$ can be written in the following form in the field-aligned coordinate:
\begin{equation}
\mathbb{P}=\begin{bmatrix} P_{\parallel} & P_{a} & P_{b}\\ P_{a} & P_{\perp} & P_c\\ P_{b} & P_c & P_{\perp}\end{bmatrix}.
\label{dp}
\end{equation}
this pressure tensor can be uniquely decomposed into a gyrotropic part $\mathbb{G}$ in the form of Eq.~(\ref{gp}) and a nongyrotropic part $\mathbb{N}$
\begin{equation}
\mathbb{N}=\begin{bmatrix} 0 & P_{a} & P_{b}\\ P_{a} & 0 & P_c\\ P_{b} & P_c& 0\end{bmatrix},
\end{equation}
such that $\mathbb{P}=\mathbb{G}+\mathbb{N}$.

Using this decomposition, Aunai {\it et.\thinspace al.}\cite{aunai13pop} proposed an agyrotropy measure
\begin{equation}
 D_{ng} \equiv  \frac{2\sqrt{\sum_{i,j} N_{ij}^2}}{tr(\mathbb{P})}
 = \frac{(8(P_a^2+P_b^2+P_c^2))^{1/2}}{P_{\parallel}+2P_{\perp}}.
\end{equation}  
An alternative is proposed by Swisdak \cite{swisdak16grl} as:
\begin{eqnarray}
 Q\equiv\frac{P_a^2+P_b^2+P_c^2}{P_{\perp}^2+2P_{\parallel}P_{\perp}}.
\end{eqnarray}
Ignoring the numeric factor, the numerator of $D_{ng}^2$ and $Q$ are the same,  but in $D_{ng}$ the denominator is the trace of the pressure tensor while in $Q$ it is a quadratic function of the diagonal elements.

The contradicting definitions of non-gyrotropy/agyrotrpy measure has caused confusion and naturally raised the question of how arbitrary one can define agyrotropy:  should the freedom of choice limited by some physical and mathematical principles? What is a good measure of agyrotropy?  
These are the issues we intend to address in this paper. 
While this study is motivated by {\it in situ} observations of the magnetic reconnection in the magnetosphere, the subject has much broader applications in plasma physics, such as turbulence in which microscopic reconnection is thought to be important. It is not the purpose of this paper to discuss if a certain method may work in specific observations or conditions. 

We first consider the basic properties a good measure of  agyrotropy should have, and based on these considerations we propose a new independent non-gyrotropic measure $AG$. We then examine $AG$ and re-examine the non-gyrotropic measures previously proposed, namely, $A\varnothing_e$,  $D_{ng}$, and  $Q$ in cases of different magnetic field alignments. We find that only $AG$ is well-behaved in all these cases.
As a demonstration of the method, we examine and compare $AG$, $Q$, $D_{ng}$ and $A\varnothing_e$ in particle-in-cell simulations of magnetic reconnection. Space observations have shown that turbulence is very important in magnetic reconnection\cite{kho16grl,ergun16grl,torbert16grl}, but its influence on agyrotropy has not been investigated previously. We analyze both turbulent and non-turbulent magnetic reconnection simulations with both force-free and Harris current sheets. We find that $A\varnothing_e$ can not properly trace the EDR and turbulent current broadening effect in force-free magnetic reconnection.
 
\section{Measuring agyrotropy}
\subsection{Basic Considerations}
\label{principle}
Measurement of any quantity generally involves comparison with some precisely defined unit value of the quantity. The definition should be unique and reflect the property investigated. Applying these basic principles to agyrotropy -- a derived quantity from pressure tensor, we must first define a quantity that describes gyrotropy, then the departure from this quantity is the measure of agyrotropy. 

We hence consider the following basic requirements for a scalar measure of agyrotropy in pressure:  
(a) The gyrotropic pressure should be uniquely defined for any given pressure tensor. Since the decomposition of  $\mathbb{P}$ into $\mathbb{G}$ and $\mathbb{N}$ is unique\cite{aunai13pop}, $\mathbb{G}$ is the unique gyrotropic tensor associated with $\mathbb{P}$;  
(b) The function that maps $\mathbb{G}$ to a scalar, i.e., $F(\mathbb{G})$ should be the same function that operates on  $\mathbb{P}$ so that $F(\mathbb{P}) - F(\mathbb{G})$ measures the departure from gyrotropy; Note that $F$ is not required to be a linear function of pressure tensor, and in general $F(\mathbb{N})=F(\mathbb{P-G}) \neq F(\mathbb{P}) - F(\mathbb{G})$. 
(c) Because gyrotropy $\mathbb{G}$  depends on the direction of local magnetic field, the scalar agyrotropy measure should reflect this dependence;  {(d) While the representation of pressure tensor $\mathbb{P}$ depends on the choice of orthogonal coordinate base $(\hat{x}_1,\hat{x}_2,\hat{x}_3)$, i.e. $P_{ij}=\mathbf{\hat{x}_i \cdot \mathbb{P} \cdot \mathbf{\hat{x}}_j}$, where $i,j=$1,2,3, the scalar function $F$ should not depend on a specific coordinate system in which $P$ and $G$ are measured, so that the agyrotropy measure is coordinate independent. The obvious choices for such scalar operators that satisfies (b) in any coordinate are invariants under spatial rotation. 
 
Based on these basic considerations, we can construct a scalar measure of agyrotropy. The simplest way to define a coordinate independent scalar operator is to use the rotational invariants of pressure tensor $\mathbb{P}$. Mathematically there are only three such invariants:  the trace $I_1$, the sum of principle minors $I_2$, and the determinant $I_3$, i.e.,
\begin{eqnarray}
I_1(\mathbb{P})& = & tr(\mathbb{P}),\\
I_2(\mathbb{P})& =& \frac{1}{2}((tr(\mathbb{P}))^2-tr(\mathbb{P}^2)),\\
I_3(\mathbb{P})& = & det(\mathbb{P}).
\end{eqnarray}
It is obvious that $I_1$ alone cannot be used to construct an agyrotropy measure since $tr(\mathbb{P})=tr(\mathbb{G})$ does not satisfy (a)--(c). $I_2$, $I_3$ are two possible independent choices.  We first introduce a new agyrotropic measure $AG$ based on $I_3$, and then revisit the previously defined agyrotropy measures $A\varnothing_e$ , $D_{ng}$ and $Q$ and compare them with $AG$. 
\subsection{A New Measure of agyrotropy }
From the discussions above we see that it is possible to define a measure of agyrotropy based on the invariant $I_3$. By definition $\mathbb{G}$ depends on the local magnetic field, and $det(\mathbb{P})-det(\mathbb{G})$ obviously satisfies (c) and (d). In addition, $\mathbb{G}$ is also unique in that it has the largest determinant among pressure tensors with the same diagonal elements. This is because:  
\begin{eqnarray}
& &  det(\mathbb{P}) -  det(\mathbb{G}) \nonumber\\  
& = &  -P_{\parallel}P_c^2-(P_a^2+P_b^2)P_{\perp}+2P_aP_bP_c,\nonumber \\
& = & -P_{\parallel}P_c^2-(P_a^2+P_b^2)P_{\perp}+ det(\mathbb{N}),\nonumber \\
 & = &-P_{\parallel}P_c^2-(P_a+P_b)^2 (P_{\perp}-P_c)/2 -(P_b-P_a)^2(P_{\perp}+P_c)/2,
\label{depart}
\end{eqnarray}
where we used the pressure tensor defined in Eq.~(\ref{dp}).  Since the principle minor of positive semidefinite $\mathbb{P}$ requires $P_{\perp}\geq P_c$, the departure $det(\mathbb{P})-det(\mathbb{G}) \leq 0$.

In theory the absolute departure $\mid det(\mathbb{P})-det(\mathbb{G})\mid$ is sufficient to describe agyrotropy for a specific $\mathbb{P}$. In practice we need to compare agyrotropy for different $\mathbb{P}$ at different locations,  or with simulations, and a measure of relative rather than the absolute departure is more useful. It is also desirable to have the relative agyrotropy to have values between 0 and 1. Thus we define the normalized agyrotropy as

\begin{eqnarray}
AG& = & \frac{\mid det(\mathbb{P})-det(\mathbb{G})\mid}{det(\mathbb{P})+det(\mathbb{G})} \nonumber \\ 
& = & \frac{\mid 4det(\mathbb{P})-P_{\parallel}(tr(\mathbb{P})-P_{\parallel})^2\mid}{4det(\mathbb{P})+P_{\parallel}(tr(\mathbb{P})-P_{\parallel})^2}.
\label{ag}
\end{eqnarray}
When one principle axis of $\mathbb{P}$ is along the local magnetic field and $det(\mathbb{P})=det(\mathbb{G})$, then $\mathbb{P}$ is gyrotropic, and $AG=0$. When none the principle axes of $\mathbb{P}$ is aligned with the local magnetic field, and the eigenvalues are small, then $det(\mathbb{P}) \ll det(\mathbb{G})$, $AG\rightarrow 1$. An example is given by Swisdak\cite{swisdak16grl} for extreme agyrotropy limit when the principle axis is not aligned with magnetic field (with one or two eigenvalues being 0) 
\begin{equation}
\mathbb{P}=\begin{bmatrix}
x & \sqrt{x} & \sqrt{x}\\ \sqrt{x} & 1 & 1\\ \sqrt{x} & 1 & 1 \end{bmatrix},
\label{px}
\end{equation}
with $x>0$. The eigenvalues are 0,0 and $x+2$. In this case $det(\mathbb{P})=0$ and the agyrotropy measure $AG=1$.

When the pressure is field aligned, i.e, with one principle axis aligned with the magnetic field and the pressure tensor is in the form of Eq.~(\ref{ag1}),  agyrotropy should depend only on the perpendicular components of the pressure because the deformation of electron orbits in such cases occur only in the perpendicular plane, limiting the momentum transport in perpendicular directions. Indeed, in this case, $AG$ becomes
\begin{equation}
AG=\frac{ (P_{\perp1}-P_{\perp2})^2}{(P_{\perp1}+P_{\perp2})^2+4P_{\perp1}P_{\perp2}}, 
\end{equation}
which is independent of $P_{\parallel}$. 

In practice, we  can also normalize agyrotropy to the range  of [0,$\infty$).  In this case, agyrotropy can be defined as
\begin{equation}
AG^{\prime} \equiv \mid 1- det(\mathbb{G})/det(\mathbb{P})\mid
\end{equation}
where $det(\mathbb{G})=P_{\parallel}(tr(\mathbb{P})-P_{\parallel})^2/4$.  For extreme agyrotropy, $AG^\prime=\infty$ due to $det(\mathbb{P})=0$. For gyrotropy, $\mathbb{P}=\mathbb{G}$ and $AG^\prime =0$. If the Pressure has a principle axis aligned with the magnetic field, then $AG^\prime=(P_{\perp1}-P_{\perp2})^2/(P_{\perp1}P_{\perp2})$, also independent of $P_\parallel$. 
\subsection{ Revisiting $Q$,  $D_{ng}$ and $A\varnothing_e$}
\label{q}
As we have discussed in the preceding section,  the invariant $I_2(\mathbb{P})=((tr(\mathbb{P}))^2-tr(\mathbb{P}^2))/2$  may also be used to define a measure of agyrotropy thus that the absolute departure from gyrotropy is $\mid I_2(\mathbb{P}) - I_2(\mathbb{G}) \mid$. It is easy to show
\begin{eqnarray}
\label{qdepart}
\nonumber
&& \mid I_2(\mathbb{P}) - I_2(\mathbb{G}) \mid\\ \nonumber
&=& \mid -\frac{1}{2}(tr(\mathbb{P}^2)-tr(\mathbb{G}^2))\mid\\ \nonumber
&=&\mid I_2(\mathbb{N})\mid \\ \nonumber
&=& \mid -P_a^2-P_b^2-P_c^2\mid\\
&=& P_a^2+P_b^2+P_c^2,
\end{eqnarray}
where we used $tr(\mathbb{P})=tr(\mathbb{G})$ and $tr(\mathbb{G}\mathbb{N})=tr(\mathbb{N}\mathbb{G})=0$.
 The absolute departure defined in Eq.(\ref{qdepart}) is a monotonic function of $P_a$, $P_b$ and $P_c$.  One way to normalize this measure is simply to divide the absolute departure by the sum of $I_2(\mathbb{P})$ and $I_2(\mathbb{G})$. Since  $\mid I_2(\mathbb{P}) - I_2(\mathbb{G}) \mid/ (I_2(\mathbb{P}) + I_2(\mathbb{G}) )=-I_2(\mathbb{N})/(I_2(\mathbb{N}) + 2I_2(\mathbb{G}) )$, the ratio is zero, i.e. gyrotropy,  if $I_2(\mathbb{N})=0$. If $I_2(\mathbb{N})\neq 0$,  for extreme agyrotropy limit shown in Eq. (\ref{px}), the maximum of the ratio is 1/3, therefore a factor of 3 is needed to ``scale" this definition to [0, 1].  Another way to normalize $I_2(\mathbb{N})$ is dividing it by $I_2(\mathbb{G})$, which yields the agyrotropy measure $Q$ first proposed by Swisdak\cite{swisdak16grl}:
\begin{eqnarray}
\nonumber Q &=& \frac{\mid I_2(\mathbb{P}) - I_2(\mathbb{G}) \mid}{I_2(\mathbb{G})}\\\nonumber
&=& -\frac{-I_2(\mathbb{N})}{I_2(\mathbb{G})}\\\nonumber
&=&\frac{P_a^2+P_b^2+P_c^2}{P_{\perp}^2+2P_{\parallel}P_{\perp}},
\end{eqnarray}
where $I_2(\mathbb{G})=\dfrac{1}{2}[(tr(\mathbb{G}))^2-tr(\mathbb{G}^2)]=P_{\perp}^2+2P_{\parallel}P_{\perp}$. Since the principle minors of positive semidefinite $\mathbb{P}$ requires $I_2(\mathbb{N})\leq I_2(\mathbb{G})$,  we have $Q\leq 1$.  Replacing $P_{\perp}$ by $(tr(\mathbb{P})-P_{\parallel})/2$ in $I_2(\mathbb{G})$, we obtain $I_2(\mathbb{G})=(tr(\mathbb{P})-P_{\parallel})(tr(\mathbb{P})+3P_{\parallel})/4$, and hence the expression in  Swisdak (2016)\cite{swisdak16grl}:
\begin{equation}
Q=1-\frac{4I_2(\mathbb{P})}{(tr(\mathbb{P})-P_{\parallel})(tr(\mathbb{P})+3P_{\parallel})}.
\end{equation}
Thus we have shown that $Q$ is in fact a $I_2$-based measure. 

When the pressure is field aligned, Q becomes 
$$Q=(P_{\perp1}-P_{\perp2})^2/((P_{\perp1}+P_{\perp2})(4P_{\parallel}+P_{\perp1}+P_{\perp2})),$$ 
which is dependent on $P_{\parallel}$, a property that is not desirable. When $P_{\parallel} \gg P_{\perp}$, $Q \rightarrow 0$ regardless of the value of $P_{\perp1}/P_{\perp2}$. This causes ambiguity, particularly between extreme agyrotropy where $P_{\perp1}/P_{\perp2} \approx 0$ or $P_{\perp2}/P_{\perp1} \approx 0$, and gyrotropy. We find this to be a generic problem for the relative agyrotropy measures defined by $I_2$ since in the numerator $I_2(\mathbb{P})-I_2(\mathbb{G})$ is independent of $P_{\parallel}$, while the denominators -- $I_2(\mathbb{P})$ or $I_2(\mathbb{G})$ or their combinations, are all dependent on $P_{\parallel}$. Since $P_{\parallel}$ in general is not constant and varies with location,  the spatial scaling of $Q$ is consequently not uniform in space. In Fig.~\ref{compare} different agyrotropy measures of field aligned pressure tensors are shown as functions of $P_{\perp 1}/P_{\perp 2}$ between 0 to 1. While $A\varnothing_e$ and $AG$ have the correct value of 1 for extreme agyrotropy, i.e, $P_{\perp 1}/P_{\perp 2}=0$,  $Q$ and $D_{ng}$ clearly decrease with the value of $P_\parallel/P_\perp$, and in both cases shown they are much smaller than 1.  Let $P_{\perp1}=0$, $P_{\perp2}=x$ and $P_{\parallel}=\alpha x$, where $x>0$ and $\alpha >1$,  we have $Q(P_{\perp 1}/P_{\perp 2}=0)=0.25/\alpha$. When $\alpha \gg 1$, we have $Q \rightarrow 0$. This completely mixes up agyrotropy with gyrotropy.   
\begin{figure}
\includegraphics[scale=0.7, trim=0 0 0 0,clip]{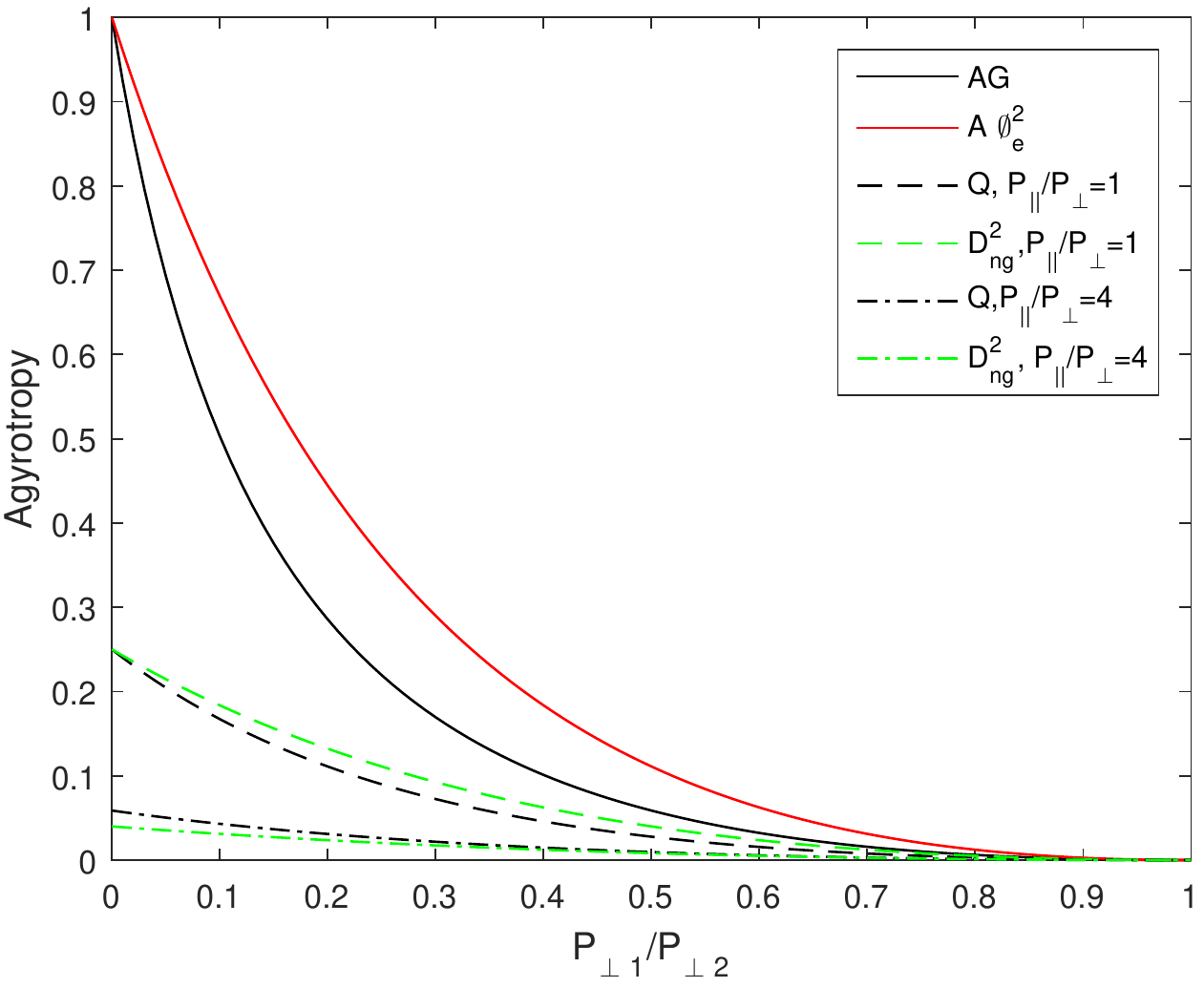}
\caption{agyrotropy measures as a function of $P_{\perp 1}/P_{\perp 2}$ when one principle axis of $\mathbb{P}$ is aligned with magnetic field. $AG$ (black solid line) and $A\varnothing_e$(red solid line) are independent of $P_{\parallel}$ while $Q$ and $D_{ng}$ show strong dependence on $P_{\parallel}$.  }
\label{compare}
\end{figure}

Now we look at $D_{ng}$. From the definition $D_{ng}=2(\sum_{ij}\mathbb{N}_{ij}^2)^{1/2}/tr(\mathbb{P})$, we have  $D_{ng}=(8 I_2(\mathbb{N}))^{1/2}/tr(\mathbb{P})$ given that $\sum_{ij}\mathbb{N}_{ij}^2=2I_2(\mathbb{N})$. It is clear that  $D_{ng}$ is also $I_2$ based like $Q$, except that it is normalized by $tr(\mathbb{P}) = I_1(\mathbb{P})$. By this definition $D_{ng}$ is not strictly a measure of relative departure from gyrotropy.  Similar to $Q$ (see Fig.\ref{compare}), $D_{ng}$ also depends on $P_{\parallel}$ when the pressure is field aligned, i.e., 
\begin{equation}
D_{ng}=\frac{(P_{\perp1}-P_{\perp2})^2}{(P_{\perp1}+P_{\perp2})(P_{\parallel}+P_{\perp1}+P_{\perp2})},
\end{equation}
causing unwanted ambiguity.  
Using $I_1$ as normalization in fact also cause problem in cases when the pressure tensor is not field aligned. To illustrate this point, let us consider the maximum value of $D_{ng}$. Since $\mathbb{P}$ being positive semidefinite  requires $I_2(\mathbb{N})\leq I_2(\mathbb{G})$, we have $D_{ng} \leq (8 I_2(\mathbb{G}))^{1/2}/tr(\mathbb{G})$. We can see that instead of being a constant, the maximum value of $D_{ng}$ is a function of $P_{\parallel}$ and $P_{\perp}$, but these two values are spatial functions. This means the scaling of $D_{ng}$ at different locations is not uniform and is also a spatial function, rendering it difficult to have meaningful comparison of agyrotropic effect at different locations when the magnetic field and plasma are nonuniform. 
\begin{figure}
\includegraphics[scale=0.3, trim=50 0 100 160,clip]{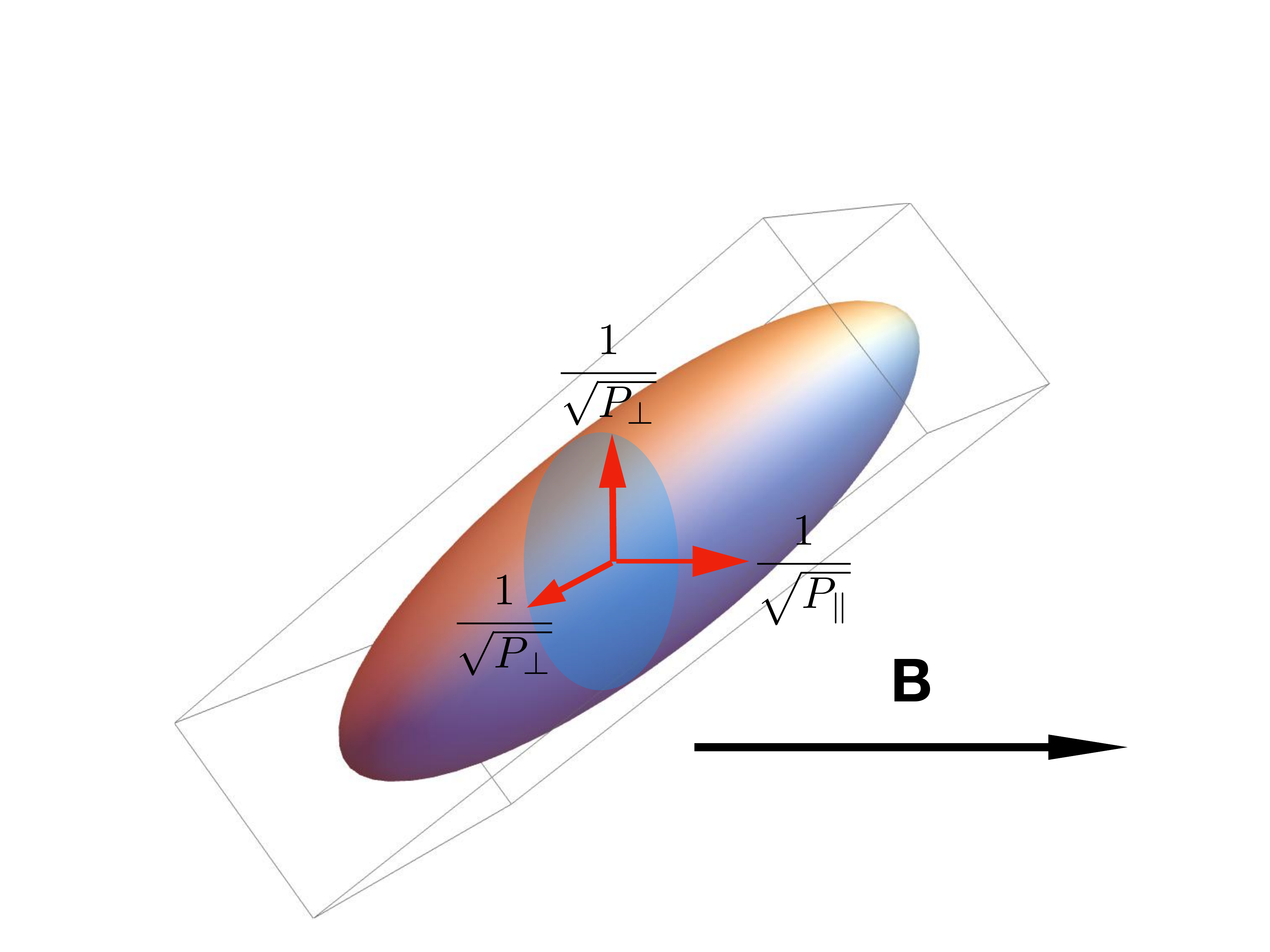}
  \caption{The ellipsoid of agyrotropic pressure tensor in a magnetic field aligned coordinate showing the principle axes of the ellipsoid tipping away from the magnetic field. Since the two perpendicular pressure component are equal, this tensor is considered under the definition $A\varnothing_e$. }
 \label{agfig}
 \end{figure}

$A\varnothing_e$ is defined in a special coordinate formed by the local magnetic field and the two eigenvectors of the perpendicular part of the pressure tensor, and the measure is independent  of $P_{12}^{\prime}$ and $P_{13}^{\prime}$ in Eq.~(\ref{aphi}).  According to  the definition of $A\varnothing_e$,  tensor 
\begin{equation}
\mathbb{P}_{s}=\begin{bmatrix} P_{\parallel} & P_{a} & P_{b}\\ P_{a} & P_{\perp} & 0\\ P_{b} & 0 & P_{\perp}\end{bmatrix}
\label{s}
\end{equation}
is gyrotropic, i.e.  $A\varnothing_e=0$ . In other words, gyrotropy corresponds to an infinite set (or equivalent class) of tensors with the same diagonal elements, thus the definition contradicts the commonly accepted  notion of gyrotropy tensor.  In the geometrical representation of the pressure tensor $\mathbb{P}_s$ (with $P_a \neq 0$ and $P_b\neq 0$) corresponds to an ellipsoid whose principle axes tip away from the magnetic field,  as shown in Fig.~\ref{agfig}. Such pressure tensors are clearly non-gyrotropic.  $A\varnothing_e=0$ simply defines a class of pressure ellipsoids whose cross sections perpendicular to the magnetic field are circular. Thus the definition of $A\varnothing_e$ can not distinguish  agyrotropic $\mathbb{P}_s$ from real gyrotropic $\mathbb{G}$. 

We further examine the condition for $A\varnothing_e=0$ to approximately correspond to gyrotropy. Without losing generality, we assume the principle axes of the pressure ellipse $\mathbb{P}_s$ only tip away from the magnetic field in the $xy$-plane in the magnetic field-aligned coordinate as shown in Fig.~\ref{agfig}. Then the axes of the ellipsoid can be found as 
\begin{gather*}
\nonumber P_{s,x}=\frac{1}{2}[(P_{\parallel}+P_{\perp})+\sqrt{(P_{\parallel}-P_{\perp})^2+4(P_a^2+P_b^2)} ],\\ \nonumber
P_{s,y}=\frac{1}{2}[(P_{\parallel}+P_{\perp})-\sqrt{(P_{\parallel}-P_{\perp})^2+4(P_a^2+P_b^2)} ], \\ \nonumber
P_{s,z}=P_{\perp}.\\ \nonumber
\end{gather*}
We can see that only when $4(P_a^2+P_b^2) \ll (P_\parallel-P_\perp)^2 $ for $\vert P_\parallel - P_\perp \vert \gg 0$, or $4(P_a^2+P_b^2) \ll (P_\parallel+P_\perp)^2 $ for $P_\parallel \sim  P_\perp$,  $\mathbb{P}_s$ reduces to the usual gyrotropic tensor.  However, satisfying these conditions are not guaranteed in processes such as magnetic reconnection; and in such cases $A\varnothing_e=0$  does not define gyrotropy correctly and would miss important contributions from electrons whose orbits are not deformed only in the perpendicular plane.  Thus  $A\varnothing_e$ in general is not a good agyrotropic measure. 

The definition of $A\varnothing_e$ is not constructed with any of the three rotational invariants in mind, and is more intuitive than other measures that applies to unconstrained pressure tensors. However, it is easy to show that $A\varnothing_e$ is actually an $I_2$-based construction. In fact, the $I_2$-based agyrotropy measure $Q$ is reduced to $A\varnothing_e$ (ignoring the factor of 2 from the original definition) when $P_\parallel = 0$. In other words,  $A\varnothing_e$ is a degenerate case of $Q$ when $P_\parallel$ can be ignored.  
 
\subsection{Comparison of $AG$, $Q$, $D_{ng}$ and $A\varnothing_e$ in Simulations of Magnetic Reconnection}
\label{p3d}
In this section we analyze three particle-in-cell (PIC) simulations performed with the \textit{p3d} code \cite{zeiler02jgr} to compare how well $AG$, $Q$, $D_{ng}$ and $A\varnothing_e$ can track or ``illuminate" electron diffusion structures of magnetic reconnection. We demonstrate in these simulations that $AG$ is a robust indicator of the EDR.  $Q$ and $D_{ng}$ also appear to track the EDR reasonably well, because while $Q$ and $D_{ng}$ are biased measures of agyrotropy, they only fail catastrophically under extreme conditions that are not met in these simulations. On the other hand,  $A\varnothing_e$ peaks well outside the EDR in both 2D and 3D force-free magnetic reconnection simulations, thus the simulations provide a concrete example to show that $A\varnothing_e$ should not be used as an EDR indicator as it was originally intended. 

\begin{figure}
\includegraphics[scale=0.65, trim=50 300 10 80,clip]{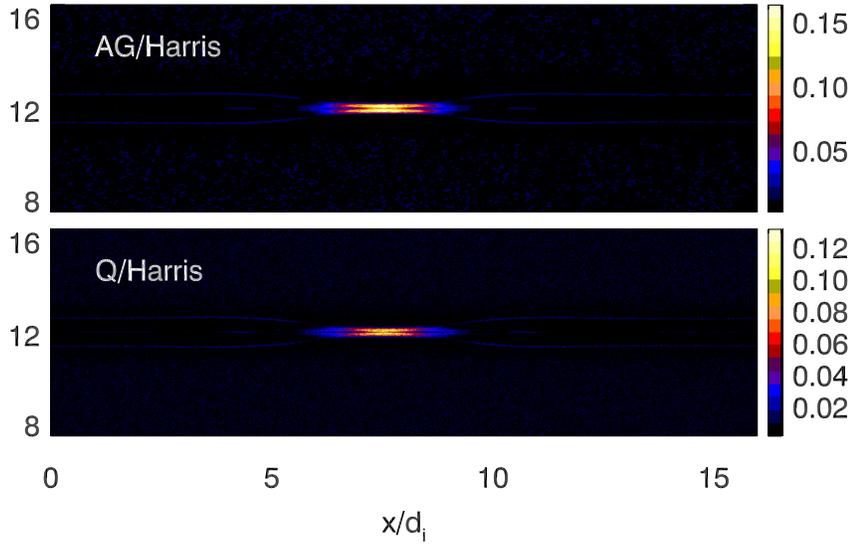}
 \caption{$AG$ and $Q$ in Harris current sheet magnetic reconnection. Both have values significantly higher in the EDR than elsewhere, but the values of $Q$ are generally smaller than $AG$. }
 \label{ag_harris}
 \end{figure}
Comparisons of different agyrotropy measures using magnetic reconnection simulations with Harris current sheet has been made previously\cite{swisdak16grl}. In the following we only use a 2D Harris current sheet reconnection simulation to demonstrate the performance of $AG$. The domain size of the simulation is $16 d_i \times 8 d_i$,  where $d_i=c/\omega_{pi,0}$ is the ion inertial length, and $\omega_{pi,0}=(4\pi n_0 e^2/m_i)^{1/2}$.  In Fig~\ref{ag_harris}, we show $AG$ and $Q$ in the Harris current sheet reconnection. It is not surprising that both measures trace the EDR in a similar fashion. The value of $Q$ is smaller than $AG$ due to the effect of $P_{\parallel}$ which we found to be larger than the perpendicular components of the pressure tensor. In Fig.~\ref{ag_harris_scudder}, $A\varnothing_e$ behaves similarly to $AG^{1/3}$ (We compare $AG^{1/3}$  with $A\varnothing_e$ because the two quantities have similar values in Harris current sheet thus offer better visual comparison). In the previous studies, it is found that $A\varnothing_e$ behaves similarly to $Q^{1/2}$ in Harris current sheet magnetic reconnection simulations with or without guide field\cite{swisdak16grl}. This is because Harris current sheet reconnection is essentially a 2D configuration even with a uniform guide magnetic field,  and in non-turbulent Harris current sheet reconnection $A\varnothing_e$ can behave relatively well in tracing current sheet and diffusion regions\cite{swisdak16grl}. $D_{ng}$ also can track these structures. 
\begin{figure}
\includegraphics[scale=0.65, trim=50 300 0 50,clip]{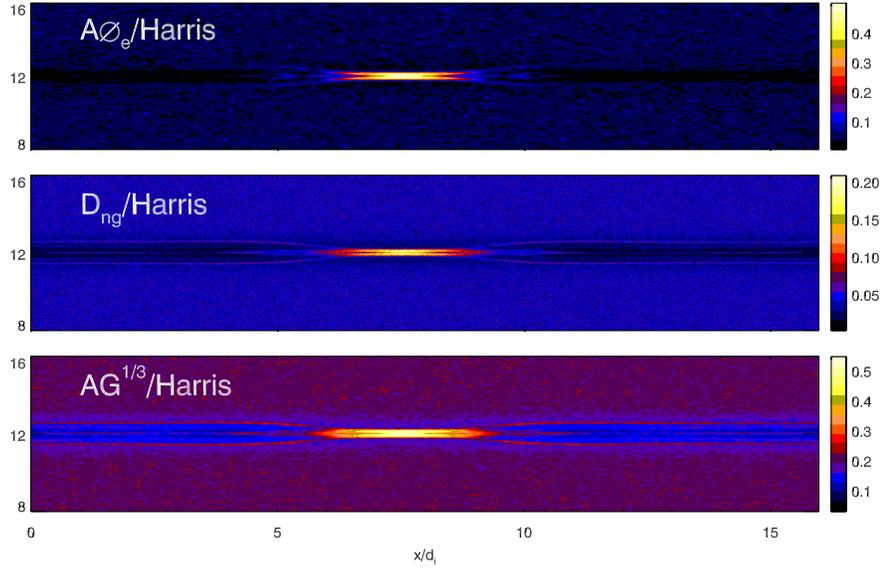}
 \caption{$AG^{1/3}$ and $A\varnothing_e$ in Harris current sheet magnetic reconnection. Both quantities clearly illuminate the EDR. }
 \label{ag_harris_scudder}
 \end{figure}
 
The initializations of the pair of force-free current sheet reconnection simulations are the same except that one is in 2D  while the other in 3D.  In the 3D simulation, strong turbulence around the x-line develops at the late stage due to the nonlinear growth of Buneman instability, while in the 2D simulation turbulence cannot develop because the Buneman instability only grows in the direction perpendicular to the reconnection plane. 
The domain size for the 2D simulation is $4 d_i\times 2 d_i$ while the 3D simulation domain is $4 d_i\times 2 d_i\times 8 d_i$. The simulation time is presented in the unit of  ion cyclotron time $\Omega_{i0}^{-1}=m_ic/eB_0$, and $n_0$ and $B_0$ are the asymptotic density and magnetic field. The guide field is $B_g=5B_0$. Both simulations have total simulation time $\Omega_{i0} t=4$. The small box simulations can demonstrate clearly the structure of the EDR. The detailed analysis of these simulations can be found in previous publications\cite{che10grl,che11nat}. 
\begin{figure}
\includegraphics[scale=0.65, trim=50 300 80 150,clip]{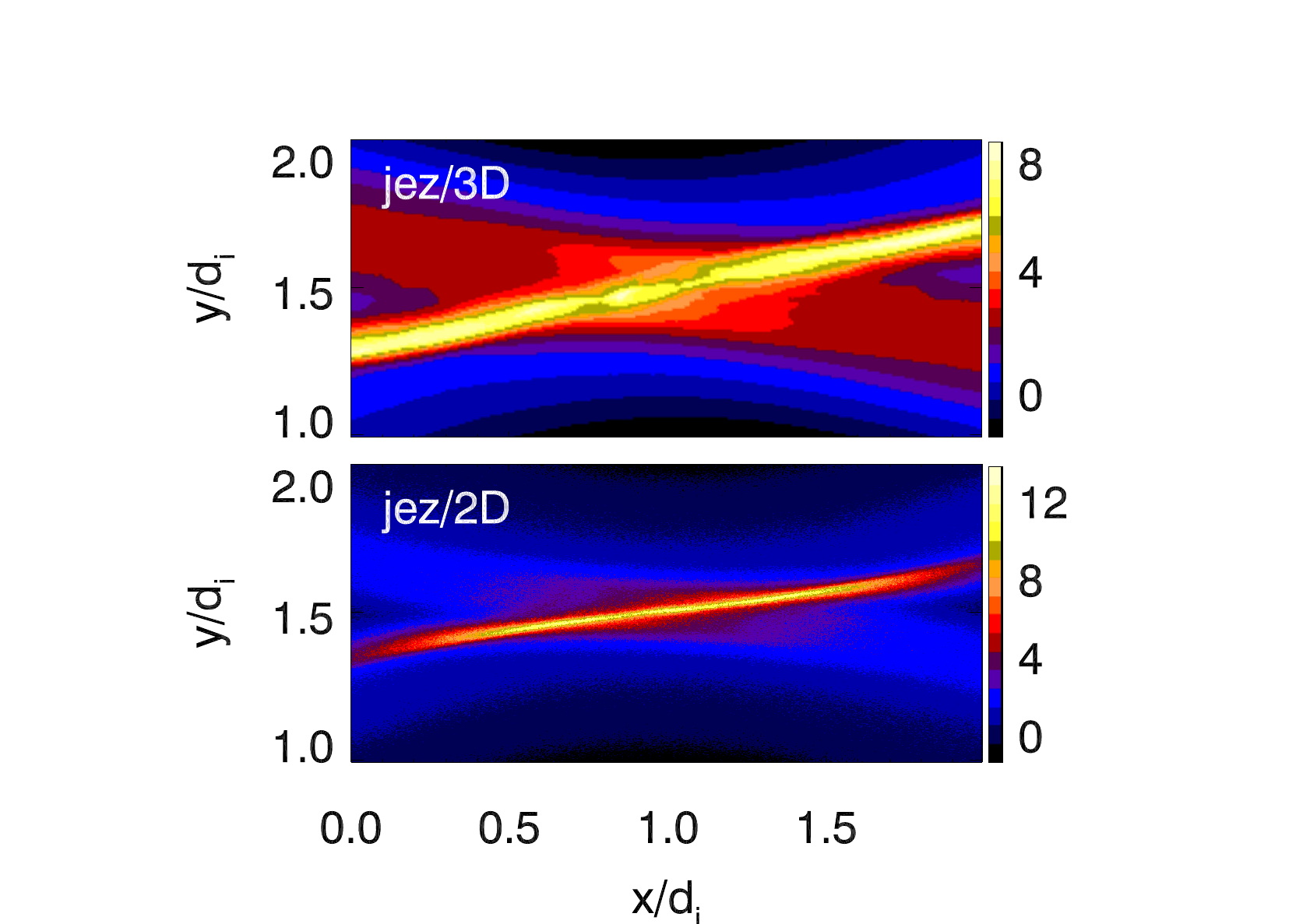}
\noindent\includegraphics[scale=0.45, trim=0 300 60 50,clip]{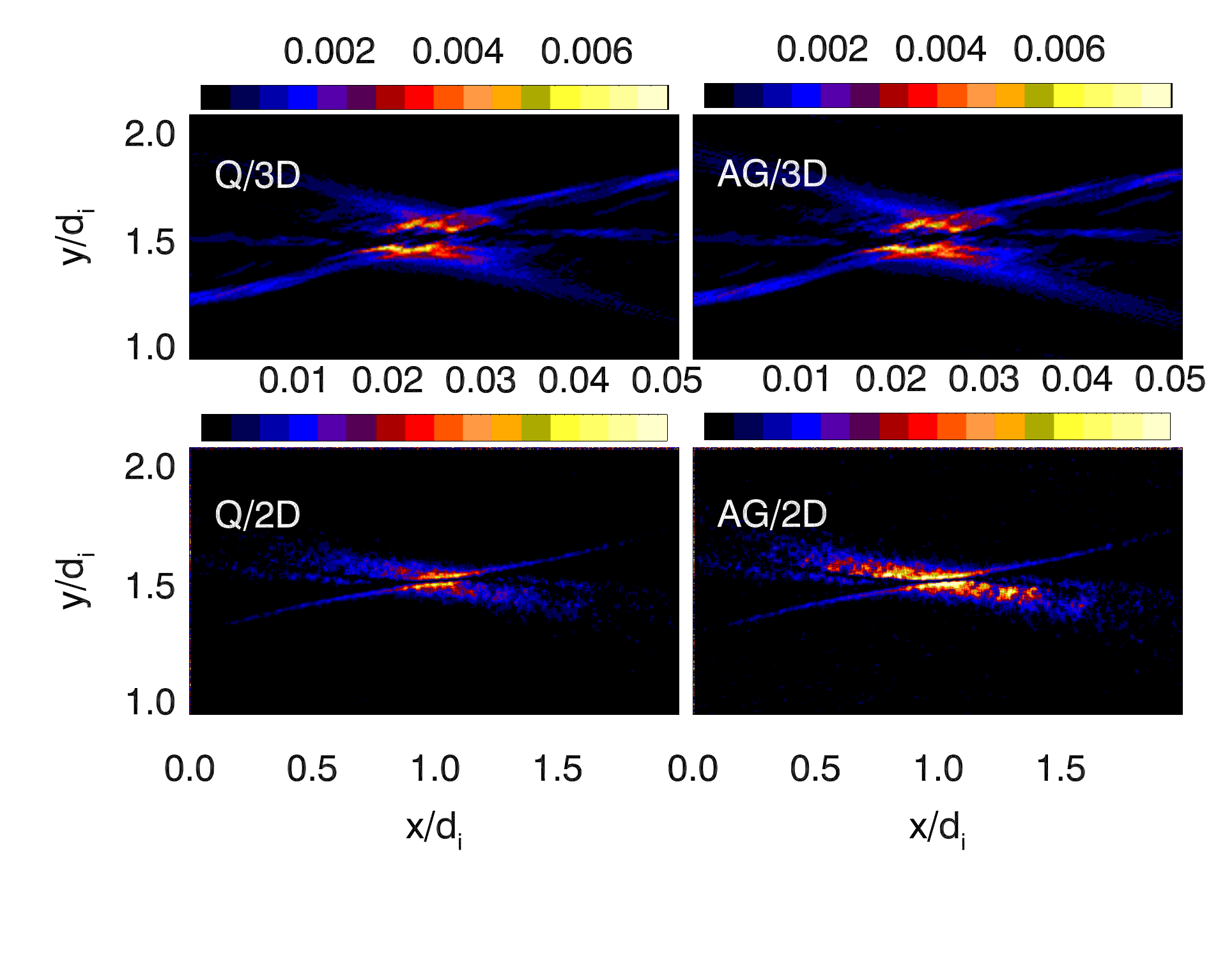}
 \caption{Electron currents and agyrotropy measures in the 2D and 3D simulations of force-free magnetic reconnection with a strong guide field (\S\ref{p3d}).  Top panels: the out-of-plane current density $j_{ez}$ at $\Omega_{i0}t=4$. Bottom panels: $AG$ and $Q$ at $\Omega_{i0}t=4$. }
 \label{jez_ag}
 \end{figure}

Force-free current sheet reconnection intrinsically is a 3D configuration in which non-uniform guide magnetic field is the largest at the current sheet. Moreover, turbulence can also cause more complex non-Larmor electron orbits. Whether the measurement can catch the turbulence effects is an important factor to evaluate the robustness of the method to measure agyrotropy. In Fig.\ref{jez_ag}, the out of plane electron current density $j_{ez}$ at $\Omega_{i0}t=4$ from both the 2D and 3D magnetic reconnection simulations are shown. The current sheet in the 3D magnetic reconnection is much broader than that in the 2D magnetic reconnection due to the turbulence scattering \cite{che11nat}. We also show both $AG$ and $Q$ in the 2D and 3D simulations in Fig.~\ref{jez_ag}.  The values of $Q$ and $AG$ are relatively small and $Q$ behaves very similarly to $AG$. Both $AG$ and $Q$ reach their maxima around the x-line and show two peaks. In the 2D simulation, the EDR is short in the x-direction and is very narrow with dimensions $\Delta x \times \Delta y \approx 0.5 d_i \times 0.1 d_i$. In the 3D simulation, the EDR becomes longer and broader with  $\Delta x \times \Delta y \approx 1 d_i\times 0.5 d_i$.   
\begin{figure}
\includegraphics[scale=0.45, trim=0 300 60 50,clip]{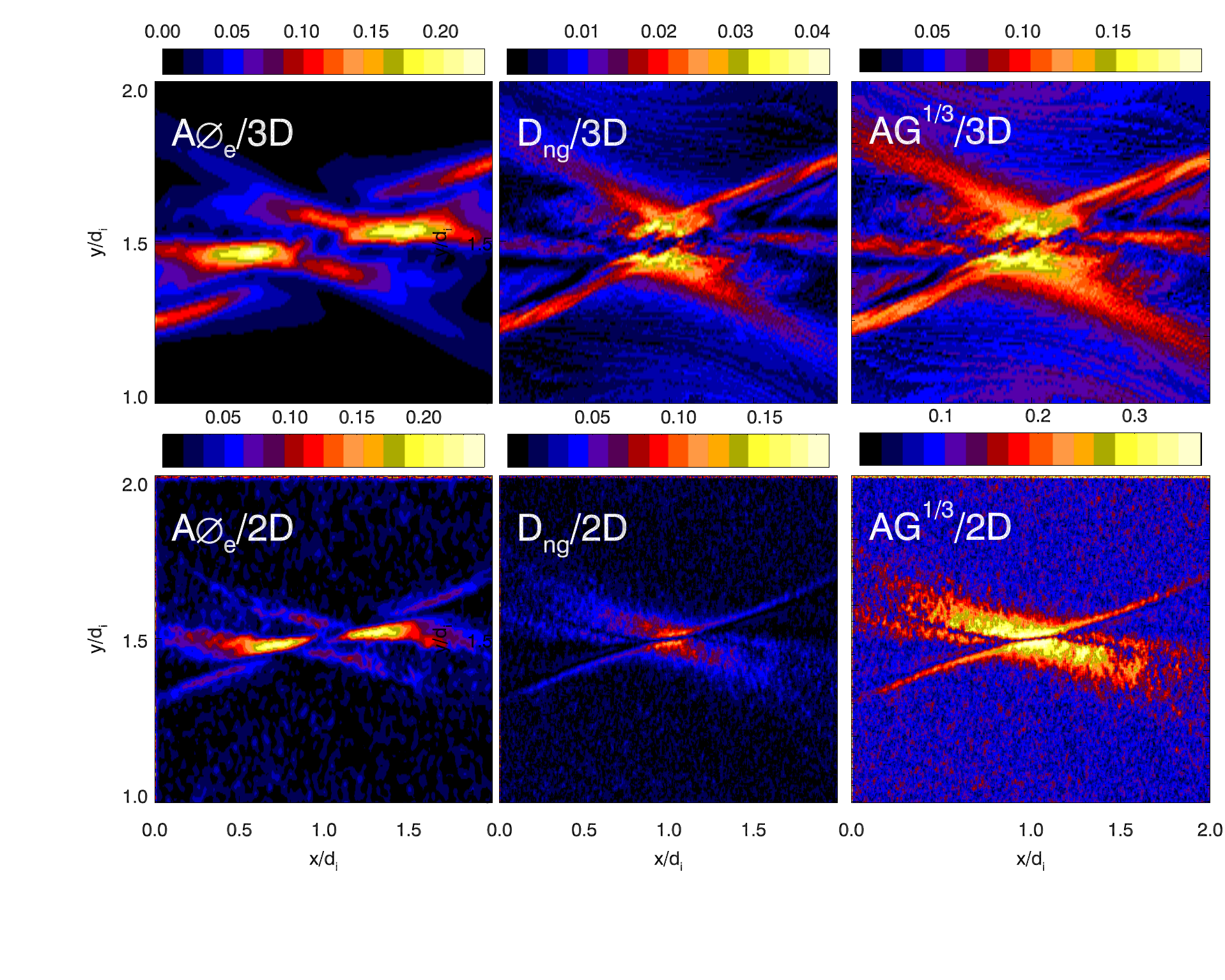}
 \caption{$A\varnothing_e$, $D_{ng}$ and $AG^{1/3}$ in force-free reconnection simulations. }
\label{ag3d_2d_scudder}
\end{figure}
 
In Fig.~\ref{ag3d_2d_scudder},  $D_{ng}$ appear similar to $AG^{1/3}$ as expected.  On the other hand, $A\varnothing_e$ peaks well outside the EDR, thus clearly misses the EDR and part of the separatrix in both 2D and 3D force-free reconnection simulations. This makes $A\varnothing_e$  an ineffective indicator of the EDR.

In 3D force-free simulation, turbulence broadens the EDR and the current sheet. The  effect is traced by $AG$ in Fig.~\ref{ag3d_2d_scudder}, demonstrating agyrotropy can be used to trace turbulence in reconnection.  Turbulence broadening and the enhancement of the current sheet are fundamentally important in magnetic reconnection because they could play important roles in fast magnetic reconnection\cite{che17pop,che11nat}. Using agyrotropy measure to diagnose turbulence has several advantages: 1) the dimensionless measurements can be compared directly between observations and simulations; 2) The stochastic motion of particles is directly associated with turbulence and thus the agyrotropy measurement is a useful indicator of turbulence effects. If we combine the agyrotropy measurement with magnetic field and other physical quantities, we may better diagnose the role of turbulence in magnetic reconnection. 
\section{Concluding remarks}
Various physical processes in magnetic reconnection and plasma turbulence  cause the electron orbits to significantly deviate from the usual guiding center behavior, leading to agyrotropy of their velocity distribution function. Quantification of agyrotropy effect on the pressure tensor has important applications in space and plasma physics. For example such measures are used to study the EDR in magnetic reconnections and in current sheets.  However, the multitude of existing measures causes confusion and raises the question of how good these measures are, how many ways agyrotropy can be defined, and what is the best way to measure agyrotropy.  In this paper we have attempted to answer these questions. 

After considering the basic properties an agyrotropy measure should have, we show that the simplest way to measure agyrotropy is to use the rotational invariants of the pressure tensor. We have ruled out any measure based on the trace of pressure tensor $I_1$. We found that all three previously defined agyrotropy measures are constructed based on $I_2$ -- the sum of the principle minors. We show that for field-aligned pressure tensor, all the $I_2$-based measures except for $A\varnothing_e$ are dependent on $P_\parallel$, which is unphysical, and the dependence causes bias and ambiguity between extreme agyrotropy and gyrotropy.  In addition, the normalization of $D_{ng}$ causes its scaling to depend on the local magnetic field and plasma properties, regardless of whether the pressure is field-aligned.  $A\varnothing_e$ is found to be the degenerate case of $Q$ when $P_\parallel = 0$.  However, instead of having a unique gyrotropic tensor for given parallel and perpendicular components, $A\varnothing_e=0$ defines a family of tensors, most of which are not gyrotropic. This leads to the measurement of the departure from gyrotropy uncertain in 3D problems where pressure tensors are generally not field aligned. In addition, the definition in general does not account for all the effects of agyrotropy on the pressure tensor. 

We introduce a new independent agyrotropy measure $AG$, which is defined based on $I_3$. We show the properties of $AG$ as well as those of other $I_2$-based agyrotropic measures in Table~\ref{tab1} and $AG$ clearly compares favorably. In this study we have examined the possible measures of the departure from gyrotropy based on all the rotation invariant operators of the pressure tensor. Our study has eliminated both $I_1$ and $I_2$-based measures, leaving us with the only $I_3$-based measure $AG$. 

Using PIC magnetic reconnection simulations, we demonstrate that $AG$ traces the EDRs and separatrices in reconnections in both Harris and force-free current sheets. Both $Q$ and $D_{ng}$ also show their highest values in the EDR. While $AG$, $Q$, and $D_{ng}$ agree qualitatively in these simulations, we must emphasize that the non-uniform scaling of $Q$ and $D_{ng}$ makes quantitative analysis difficult. In space observations the magnetic field and plasma are highly nonuniform, to use agyrotropy quantitatively in analysis of data, a uniform scaling is essential. On the other hand, while $A\varnothing_e$ can trace the EDR in Harris reconnection, it fails to trace the EDR and part of the separatrix in force-free reconnection simulations. This demonstrates the fundamental difference between the electron dynamics in Harris and force-free current sheets.  These simulations highlight the importance of accounting for all effects of electron agyrotropy on the pressure tensor when defining a measure.  
\begin{table}[ht]
\caption{Comparison of Different agyrotropy Measures}
\begin{ruledtabular}
\begin{tabular}{ccccc}
 &$AG$ &$Q$&$D_{ng}$&$A\varnothing_e$ \\
\hline
Rotation Invariant & $I_3$ & $I_2$ & $I_2$ & $I_2$ \\
Unique Gyrotropy Defined & Y & Y & Y & N \\
Field aligned measure independant of $P_{\parallel}$ & Y & N & N & Y\\
Uniform Scaling & Y & N\footnote{The scaling depends on $P_{\parallel}$ when the tensor is field aligned.} & N & Y \\
Trace EDR in Harris reconnection & Y & Y & Y & Y \\
Trace EDR in Force-free reconnection & Y & Y & Y & N\\
\end{tabular}
\end{ruledtabular}
\label{tab1}
\end{table}

\begin{acknowledgments}
We thank the anonymous referee whose criticisms and suggestions helped to improve the paper.  We thank M. Swisdak for his very helpful discussion and critical comments on the manuscript. We also thank W. Daughton for his comments. We thank Kuznetsova for her helpful discussion on the construction of agyrotropy. We thank also the UCLA MMS team for the helpful discussion. We thank the participants of the Second MMS community Science Workshop (2017) who attended the first presentation of this work and kindly provided comments and feedback. HC is partially supported by MMS project, and by NASA grant No.NNX17AI19G. The simulations and analysis were carried out at the NASA Advanced Supercomputing facility at Ames Research Center under NASA High-End Computing Program.
\end{acknowledgments}

%

\end{document}